\documentclass[journal]{IEEEtran}

\usepackage{bm}
\usepackage{braket}
\usepackage[T1]{fontenc} 
\usepackage{float}
\usepackage{amsmath,amsfonts,amssymb,eqnarray}  
\usepackage{color, soul}
\usepackage{graphicx}  
\usepackage{dblfloatfix} 
\usepackage{array,booktabs}
\usepackage[pdftex,hypertexnames=false]{hyperref} 
\usepackage[affil-it]{authblk}
\usepackage[font={small}]{caption}
\usepackage{cite}
\usepackage[left=2cm,right=2cm,top=1.6cm]{geometry}
\usepackage{wrapfig}
\usepackage{braket}

\begin{document}

\title{Stable transmission of high-dimensional quantum states over a 2 km multicore fiber}

\author{Beatrice~Da~Lio$^{\dagger}$, Davide~Bacco*, Daniele~Cozzolino, Nicola~Biagi, Tummas~Napoleon~Arge, Emil~Larsen,  Karsten~Rottwitt, Yunhong~Ding, Alessandro~Zavatta, and~Leif~Katsuo~Oxenl{\o}we
\thanks{$^{\dagger}$bdali@fotonik.dtu.dk; *dabac@fotonik.dtu.dk}

\thanks{Manuscript received October 1, 2019; revised XXX, 2019.

This work is supported by the Centre of Excellence SPOC - Silicon Photonics for Optical Communications (ref DNRF123), by the People Programme (Marie Curie Actions) of the European Union's Seventh Framework Programme (FP7/2007-2013) under REA grant agreement n$^\circ$ $609405$ (COFUNDPostdocDTU), by the VILLUM FONDEN, QUANPIC (ref. 00025298)) and by the NATO Science for Peace and Security program  under grant G5485.

B. Da Lio, D. Bacco, D. Cozzolino, T. N. Arge, E. Larsen, K. Rottwitt, Y. Ding and L. K. Oxenl{\o}we are with Centre of Excellence SPOC, DTU Fotonik, Technical University of Denmark.

N. Biagi and A. Zavatta are with CNR-INO, Istituto Nazionale di Ottica, Firenze, Italy.}}

\markboth{Journal of Selected Topics in Quantum Electronics}%
{B. Da Lio \MakeLowercase{\textit{et al.}}: Stable transmission of high-dimensional quantum states over a 2 km multicore fiber}

\maketitle

\begin{abstract}
High-dimensional quantum states have already settled their advantages in different quantum technology applications. However, their reliable transmission in fiber links remains an open challenge that must be addressed to boost their application, \textit{e.g.} in the future quantum internet.
Here, we prove how path encoded high-dimensional quantum states can be reliably transmitted over a 2 km long multicore fiber, taking advantage of a phase-locked loop system guaranteeing a stable interferometric detection.
\end{abstract}


\section{Introduction}
\IEEEPARstart{Q}{uantum} communication, the art of transporting quantum states from one place to another, is currently a niche technology, yet to be widely deployed. One of its main goals is the quantum internet, which, despite the outstanding benefits, remains unreachable until quantum communication systems can improve in terms of rate, robustness and integration with current telecommunication infrastructures~\cite{Wenher2018}. 
In this scenario, \textit{qudits}, \textit{i.e.} quantum states spanning a \textit{d}-dimensional Hilbert space, may help addressing such issues thanks to the unique advantages they can disclose. The first and rather clear benefit, is the higher information capacity. For instance, a high-dimensional state with \textit{d}=4 can encode 2 classical bits of information. At the same time, high-dimensional quantum states are intrinsically more robust to both environmental noise and errors due to eavesdropping attacks~\cite{HiDReview,sheridan2010security,ecker2019entanglement}.

To expand the Hilbert space, \textit{i.e.} increase the dimensionality of the quantum states, different photonic degrees of freedom can be used and even combined, such as the orbital angular momentum of light~\cite{Cozzolino2018,Cozzolino2019,giordani2019experimental}, time-energy/time-bin~\cite{Islam2017,Steinlechner2017,Martin2017}, path~\cite{Wang2018,Krennpath2017,Dellantonio2018,Bacco2017} or frequency encoding~\cite{Kues2017,Jin_2016}. Each degree of freedom offers different advantages in terms of stability, control and scalability, while facing different problems~\cite{HiDReview}. For instance, the path encoding technique developed on a photonic integrated circuit offers advantages due to the intrinsic scalability and stability of the platform, which benefits from a highly repeatable production process as well~\cite{Wang2018,Ding2017}.
One of the main challenges of this approach is the reliable transmission of such states: a possibility is to couple each path to different fibers, but superposition states would be impaired by the different phase drifts. Instead, using different cores of a multicore fiber improve the quality of the transmission, as the relative phase drifts of each core are strongly suppressed since they are placed within the same cladding~\cite{xavier2019quantum,Canas2017}.
Multicore fibers propagation losses are similar to those of standard single mode fibers, and their inter-core cross-talk is low enough to ensure the reliable transmission of quantum states~\cite{hayashi2012,hayashi2017,Sasaki2017}.
Indeed, these fibers have already been investigated as a mean for high-dimensional quantum communication, however limitations in terms of stability affected the achievable distance~\cite{Ding2017,Canas2017,lee2017experimental}.

Here, we investigate the phase stability of path encoded qudits propagated in a 2 km long multicore fiber using a phase-locked loop system. This stabilization system improves both the fidelity propagation and the reachable distance of previous works, overcoming the existing limitations.

\section{Previous works}
Two simultaneous experiments, one performed by Ca\~{n}as \textit{et al.} and the other by our group, have already investigated high-dimensional quantum communication exploiting path encoded qudits over the cores of a multicore fiber~\cite{Ding2017,Canas2017}. In both works, quantum states spanning a four-dimensional Hilbert space were generated, transmitted over a multicore fiber and detected with fidelities that allowed for successful quantum key distribution applications of such systems. A third work, by Lee \textit{et al.}~\cite{lee2017experimental}, makes use of two multicore fibers to distribute four-dimensional entangled photons pairs that are path encoded over the cores of the fibers. Single photons were generated through spontaneous parametric down-conversion, and analysed, after transmission, with projective measurements using spatial light modulators and spatial mode filtering by single mode fibers. The multicore fibers used were only 30 cm long, hence phase drifts did not appear to be a major limitation. However, based on the results of the other two previous experiments and this work, any entanglement distribution setup would also require phase stabilisation when implemented on longer distance. 

In our previous work~\cite{Ding2017}, both transmitter and receiver were integrated on silicon photonics circuits that coupled light to and from the fiber cores directly, through apodized grating couplers positioned in such a way they correspond to the fiber cores position within the cladding. The fiber used in the demonstration is a 3 m long 7-core multicore fiber. Weak coherent pulses are injected into the transmitter chip where the quantum states are prepared using cascaded thermally tunable Mach-Zehnder interferometers, which split or redirect the pulses into different paths, and phase shifters to apply the required phase difference among the different paths. The modulation speed achieved in this work is actually limited to 5 kHz by the thermal dynamics of the heaters. However, the achievable extinction ratio and the good stability provide a good platform for the purposes of the experiment. On the transmitter chip there are also variable optical attenuators that are used to finely tune the power of the light, in order to increase or decrease the average photon number per pulse. The prepared quantum states belongs to three mutually unbiased bases and can be expressed as follows (the normalization factor $1/\sqrt{2}$ is omitted):
\begin{equation}
    \begin{pmatrix} \ket{A}+\ket{B}\\
    \ket{A}-\ket{B}\\
    \ket{C}+\ket{D}\\
    \ket{C}-\ket{D}
    \end{pmatrix},
    \:
    \begin{pmatrix} \ket{A}+\ket{C}\\
    \ket{A}-\ket{C}\\
    \ket{B}+\ket{D}\\
    \ket{B}-\ket{D}
    \end{pmatrix},
    \:
    \begin{pmatrix} \ket{A}+\ket{D}\\
    \ket{A}-\ket{D}\\
    \ket{B}+\ket{C}\\
    \ket{B}-\ket{C}
    \end{pmatrix},
    \label{eq_statesDing}
\end{equation}
where the states $\ket{A}$, $\ket{B}$, $\ket{C}$ and $\ket{D}$ are related to four individual cores of the fiber. At receiver side then, projection measurements in one of the bases are carried out tuning accordingly the Mach-Zehnder interferometers in order to create interference among the desired cores.
The experimental results reported show very high state preparation fidelities, of (97.3$\pm$0.6)\%, (98.1$\pm$0.6)\% and (97.5$\pm$0.6)\% for each basis in eq.~\eqref{eq_statesDing} respectively. These data were obtained by sending one state at a time with a repetition rate of 10 kHz, a mean photon number per pulse lower than 0.1 and during a 30 s acquisition window. Furthermore, we performed a quantum key distribution protocol with decoy state technique: the states were randomly chosen and prepared in a real-time fashion, sent through the multicore fiber and projected on a randomly chosen basis as well. With a repetition rate of 5 kHz, average photon numbers of 0.276 photon/pulse, 0.153 photon/pulse and the vacuum state and the use of InGaAs single photon avalanche photodiodes, the system showed stable overall quantum bit error rate of 13\% for a period of over 10 min, allowing for a final key generation rate of approximately 6.57$\times$10$^{-3}$ bit/pulse, corresponding to 33 bit/s.

The demonstration by Ca\~{n}as \textit{et al.}~\cite{Canas2017} was carried out using a 4-core 300 m long multicore fiber. The quantum states were encoded in weak coherent pulses using a combination of bulk optics and a 10 m long piece of a multicore fiber similar to the one later used as communication channel. Indeed, the pulses were coupled using a lens to all four cores of this shorter fiber, so that all the cores were equally illuminated. In such a way, they were able to create a coherent superposition state over all the four cores simultaneously. This obtained state (at the output of the short multicore fiber) was then imaged to a deformable mirror using a second lens, such that light coming from each cores hit different mirrors, whose longitudinal position can be set independently from the others. Hence, different offsets defined the different phase shifts between the cores required for generating the different quantum states. Moreover, different phase drifts accumulated along the cores could also be compensated. The states generated in the experiments were part of two mutually unbiased bases, expressed as follows (the normalization factor $1/2$ is omitted): 
\begin{equation}
    \begin{pmatrix} \ket{A}+\ket{B}+\ket{C}+\ket{D}\\
    \ket{A}-\ket{B}+\ket{C}-\ket{D}\\
    \ket{A}+\ket{B}-\ket{C}-\ket{D}\\
    \ket{A}-\ket{B}-\ket{C}+\ket{D}
    \end{pmatrix},
    \:
    \begin{pmatrix} \ket{A}+\ket{B}+\ket{C}-\ket{D}\\
    \ket{A}+\ket{B}-\ket{C}+\ket{D}\\
    \ket{A}-\ket{B}+\ket{C}+\ket{D}\\
    -\ket{A}+\ket{B}+\ket{C}+\ket{D}
    \end{pmatrix},
\end{equation}
where we use the same notation as above. The prepared quantum states were then coupled to the longer multicore fiber for the actual transmission, using another lens. At receiver side, each pulse was detected with a similar setup: a lens and a second deformable mirror were used to project into one of the possible states, directing it to one single photon detector. With a link length of 300 m, independent phase drifts between cores starts to already be observable and deteriorate the communication stability. To address this problem, the authors conceived a control system that checks whether the reference frames of transmitter and receiver are still aligned every 30 s, and, in case of negative result, a control routine that looks for the new reference is initialized. During this procedure, the quantum communication must be interrupted until a new reference point is found.
This system demonstrated that high-dimensional quantum states can be faithfully propagated over a multicore fiber with an average fidelity of (92.05$\pm$0.03)\% during 1 hour of acquisition. Also in this work, a quantum key distribution protocol was implemented, using a decoy state technique with average photon numbers of 0.2 photon/pulse, 0.1 photon/pulse and vacuum state, repetition rate of 1 kHz, gated InGaAs single photon avalanche detectors, which allowed for a stable measured quantum bit error rate of 10.25\% over more than 20 hours. The final obtained secret key rate was 4.31$\times$10$^{-6}$ bit/pulse, which corresponds to the generation of approximately 15 bit/hour.

\begin{figure*}[!ht]
\centering
\includegraphics[width=.9\textwidth]{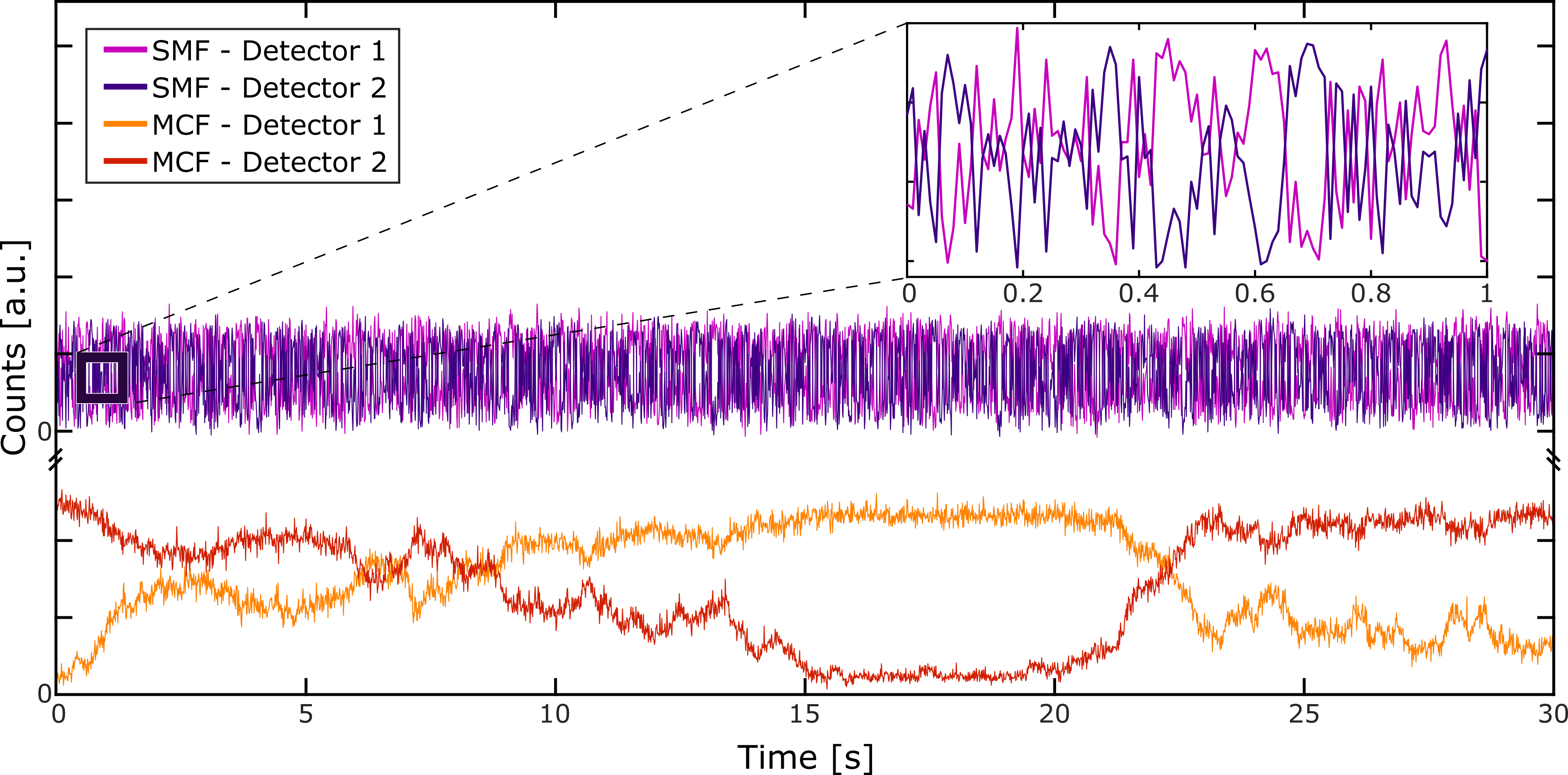}
\caption{\textbf{Stability comparison of a multicore fiber and two identical single mode fiber.} Outputs of a balanced interferometer with 2 km long arms consisting of: two single mode fibers (purple and pink traces), or two cores in the same multicore fiber (red and orange traces). The inset shows the phase drifts between the two single mode fibers over one second. Counts are collected by two single photon detectors integrated over 10 ms during an overall 30 s time window (y-axis with arbitrary units).}
\label{fig_int}
\end{figure*}

Summing up, both experiments successfully demonstrated quantum communication of high-dimensional states path encoded in the cores of a multicore fiber. Our previous work exploited a 3 m long 7-core fiber integrating both transmitter and receiver on photonic circuits, envisaging a future deployment in an integrated quantum network scenario. The work by  Ca\~{n}as \textit{et al.} instead, using a bulk-optics setup, demonstrated that transmission was possible even over 300 m long multicore fiber, monitoring the coherence with a separate control system.
However, both works still suffer from the rather short link distances. Increasing the length of the communication channel in these scenarios appears to rapidly affect the phase coherence of superposition states, an issue we address and solve in the following sections of the manuscript.

\section{Addressing the stability problem}
As already introduced in the last paragraph, the main issue of previous works appears to be the phase stability between cores, even at short link distances, which hinders the scalability of such high-dimensional quantum communication systems. However, the motivation of using a multicore fiber for the transmission of path encoded quantum states is still very strong: {\color{black} when no stabilisation system is implemented, }the phase drifts experienced by light propagating in different cores of the same multicore fiber are much slower than those experienced in separate single mode fibers. This is experimentally demonstrated in Fig.~\ref{fig_int}, where we reported in the pink and purple traces the measured detection events of two outputs of a balanced interferometer made by two 2 km long standard single mode fibers (one per arm). Instead, the orange and red traces show the two outputs of an interferometer where the arms are two cores of a 2 km long multicore fiber (a description of the fiber is given in the next section). Direct light from a continuous wave laser emitting at 1550 nm was attenuated, to not saturate the single photon detectors, and sent to the interferometers to record the interference. The collected counts from the detectors were then integrated over a 10 ms time window during a continuous period of 30 s. {\color{black} In both scenarios, the experiments took place in a lab environment having a controlled temperature, but no further temperature stabilization was implemented. We assume the temperature drifts experienced in the two cases are similar.} As can be seen from Fig.~\ref{fig_int}, the phase drifts among cores in a multicore fiber of 2 km length happen on a second time-scale, while those experienced between two independent single mode fibers take place on a millisecond time-scale.

Nonetheless, to faithfully transmit quantum states on multicore fibers, a stabilization system, tracking and compensating for phase drifts, is still required. Such a system was implemented on an electronic board, which uses as reference signal ($RS$) the detected events from a single photon detector from one output of the interferometer, giving as output an electric signal driving a phase actuator~\cite{biagi2018entangling}. The locking algorithm implemented on the board is constituted by two main parts: the \textit{scanning} part and the \textit{active locking} part. In the former, the reference signal is measured by an analog-to-digital converter of an ADUC7020 micro-controller, while a linear voltage ramp is applied to the phase actuator (in our setup a phase shifter) to produce a phase scan of 2$\pi$. In this way, it is possible to identify the maximum ($M$) and the minimum ($m$) values of the reference signal. By knowing these two parameters, it is possible to determine the value of the reference signal corresponding to a precise phase value ($\tilde{\varphi}$) according to the equation:
\begin{equation}
RS(\tilde{\varphi})=\frac{M+m}{2} + \frac{M-m}{2}\mathrm{cos}(\tilde{\varphi})
\label{ppl}    
\end{equation}
The active locking part of the algorithm is a digital implementation of a proportional-integral-derivative controller, in which the value $RS(\tilde{\varphi})$ is used as set point, \textit{i.e.} as locking point. In this process, the reference signal is continuously monitored and the voltage applied to the phase actuator is changed to keep the reference signal as close as possible to the set point $RS(\tilde{\varphi})$.

{\color{black} By running such a device at 3.5 kHz} and integrating it in our quantum communication setup, we were able to demonstrate stable interference of superposition states over long acquisition periods on a 2 km long link.

\section{Results}
\begin{figure}[!ht]
\centering
\includegraphics[width=.45\textwidth]{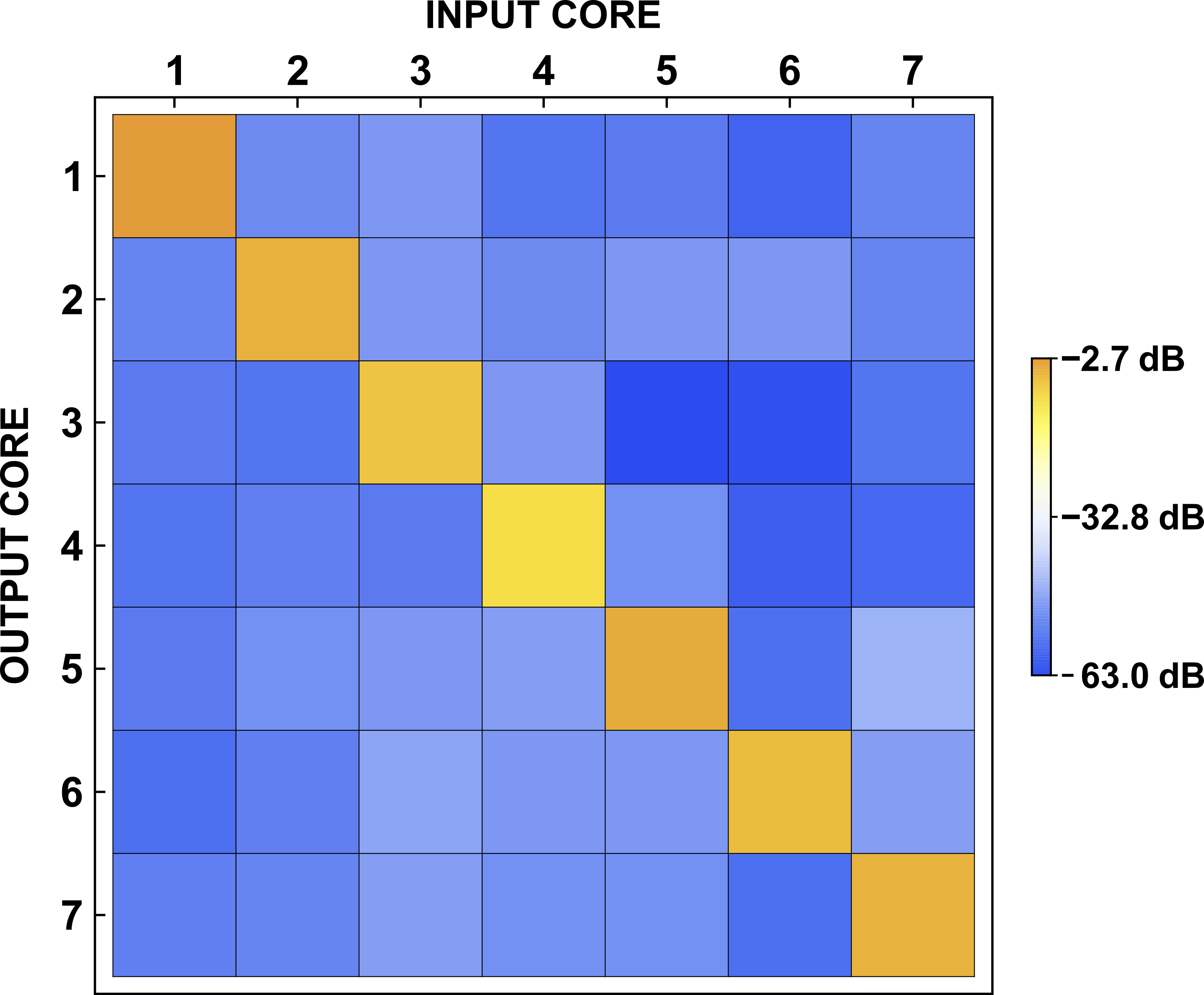}
\caption{\textbf{Measured loss per core and inter-core cross-talk.} The matrix shows in the main diagonal the losses of each core. The other elements represent the cross-talk terms.}
\label{fig_loss}
\end{figure}

The multicore fiber used as quantum communication channel in this work is a 2 km long seven-core fiber from OFS-Fitel. {\color{black}In this particular fiber, each core has a diameter of 9 $\mu$m, the core distance is 46.8 $\mu$m and the overall cladding diameter is 186.8 $\mu$m. It is a homogeneous multicore fiber, meaning that the nominal refractive index is the same in every core.} Fan-in/fan-out devices couple the light to/from each core to a different single mode fiber, to enable easier connections with the transmitter and receiver apparatuses. Losses in each core were characterized and are reported in Fig.~\ref{fig_loss}, along with the measured cross-talk. Inter-core cross-talk represents the amount of light that is coupled along the fiber from one core to another. For quantum communication purposes, the lower the cross-talk is the higher the fidelity of the transmitted states. All values of cross-talk obtained in this specific fiber are below -46 dB. Concerning the losses, the cores show values ranging from -2.7 dB to -10 dB. The difference is usually given by the fan-in/fan-out devices, rather than to the multicore fiber itself. Since the cross-talk characteristics are similar for every core, we chose the four cores that exhibit lower loss, \textit{i.e.} cores number 1, 2, 5 and 7, to encode the quantum states.
The implemented quantum states lie in a four-dimensional Hilbert space spanned by the fiber cores, as the ones realized in the works discussed above. Similarly to the chip-to-chip experiment~\cite{Ding2017}, all the states are encoded as a superposition of two cores, which allows us to be able to prepare states belonging to two mutually unbiased basis, $\mathcal{M}_0$ and $\mathcal{M}_1$, having to stabilize only one interference per state. The states can then be expressed as follows:
\begin{equation}
    \mathcal{M}_0 = \frac{1}{\sqrt{2}} \begin{pmatrix} \ket{1}+\ket{5}\\
    \ket{1}-\ket{5}\\
    \ket{2}+\ket{7}\\
    \ket{2}-\ket{7}
    \end{pmatrix}
    \:
    \mathcal{M}_1 = \frac{1}{\sqrt{2}} \begin{pmatrix} \ket{1}+\ket{7}\\
    \ket{1}-\ket{7}\\
    \ket{2}+\ket{5}\\
    \ket{2}-\ket{5}
    \end{pmatrix}
    \label{eq_states}
\end{equation}
In the following, we will refer to the states belonging to the $\mathcal{M}_0$ basis as the $\ket{\psi_i}$ states, where the subscript $i=1,\dots,4$ defines the states in the order given in eq.~\eqref{eq_states}, and similarly as the $\ket{\phi_i}$ states those belonging to the $\mathcal{M}_1$ basis.
In our setup, we prepare one of these states at a time, but at receiver side the projection is carried out over the whole set of states in that basis, which requires four single photon detectors. The phase-locked loop board that stabilizes interference is also used to switch between the 0 or $\pi$ phase difference that identifies two states sent through the same couple of cores.
\begin{figure*}[!t]
\centering
\includegraphics[width=.9\textwidth]{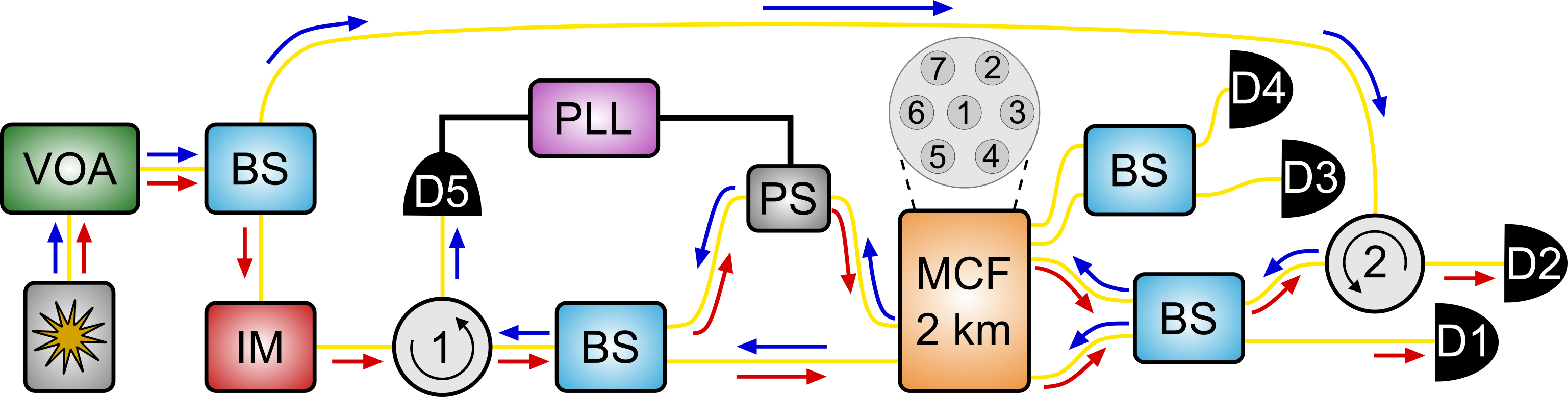}
\caption{\textbf{Schematic of the experimental setup.} Red arrow: quantum signal (1550 nm) direction of propagation; Blue arrow: control signal (1550 nm) direction of propagation; VOA: variable optical attenuator; BS: beam splitter; IM: intensity modulator; $\circlearrowleft$: circulator (arrows indicate the direction in which light is routed); PLL: phase-locked loop; PS: phase shifter; MCF: multicore fiber (dimensions of cores are exaggerated for clarity); D1, D2, D3, D4: single photon avalanche detectors used for the quantum signal; D5: single photon avalanche detector used for the control signal.}
\label{fig_setup}
\end{figure*}

The experimental setup implemented is reported in Fig.~\ref{fig_setup}. At transmitter side, a continuous wave laser emitting at 1550 nm is attenuated to the quantum regime by a variable optical attenuator (VOA). The light is then divided by a beam splitter (BS) into a co-propagating beam (red, from left to right) and a counter-propagating beam (blue, from right to left). The former is used for quantum communication and the latter as a control signal to stabilize interference between cores. In the quantum channel, pulses are carved using two cascaded intensity modulators (IM, for simplicity only one is shown in Fig.~\ref{fig_setup}), enhancing the extinction ratio, with 600 MHz repetition rate and 150 ps of optical full-width half maximum. The quantum states to be transmitted, as discussed, require a superposition among two cores of the multicore fiber (MCF) at a time: hence, the train of pulses is split by a beam splitter whose two outputs are connected to two cores of the fiber. Before the transmission, a fiber phase shifter (PS) changes the phase in one of the two arms such that a stable interference is possible at the receiver side. Such a device is driven by the phase-locked loop (PLL) board described in the previous section, and has low insertion loss (less than 0.1 dB) thanks to an all-fiber design.

The prepared quantum states can be sent through the multicore fiber channel, consisting of the multicore fiber itself with fan-in/fan-out devices and time of arrival compensation fiber patches and delay lines. These are required because the cores in a multicore fiber have slightly different {\color{black} optical path lengths: in the case of homogeneous multicore fibers, the average skew (spread in propagation time) is of the order of 100 ps/km~\cite{puttnam2019characteristics}}. Moreover, the main relative delay contribution is given by the fan-in/fan-out devices. Finally, the presence of the phase shifter in only one of the two arms must also be compensated. At receiver side, a perfect time overlap of the pulses is required to have good interference, \textit{i.e.} a high fidelity of the transmitted quantum states. Furthermore, losses must also be even to have better interference, however, as discussed above, the multicore fiber with its fan-in/fan-out devices shows different losses for different cores. The patch-cord fibers, delay lines and connector we added for compensating the time of arrival makes it such that we have a maximum channel loss of 7 dB in one of the cores. Thus, we further attenuate the other cores such that all possible paths exhibit the same loss.

Hence, the light from the two cores is directed to the final beam splitter, where interference takes place: ideally, the pulse is only detected in one of the two outputs, depending on the phase shift Alice prepared, either in detector D1 or in detector D2. Detectors D3 and D4 project onto the other two states of the basis, \textit{i.e.} the two outputs of a beam splitter connecting the other two cores. In our implementation, we do not prepare all the states in a real-time fashion, thus these last two detectors are only measuring the inter-core cross-talk. However, this always resulted to be lower than the detectors dark counts. The detectors we used are InGaAs avalanche photodiodes {\color{black} (not gated)}: parameters were set to 20\% efficiency and 20 $\mu$s dead time, giving approximately 500 Hz of dark counts.

\begin{figure}[!ht]
\centering
\includegraphics[width=.48\textwidth]{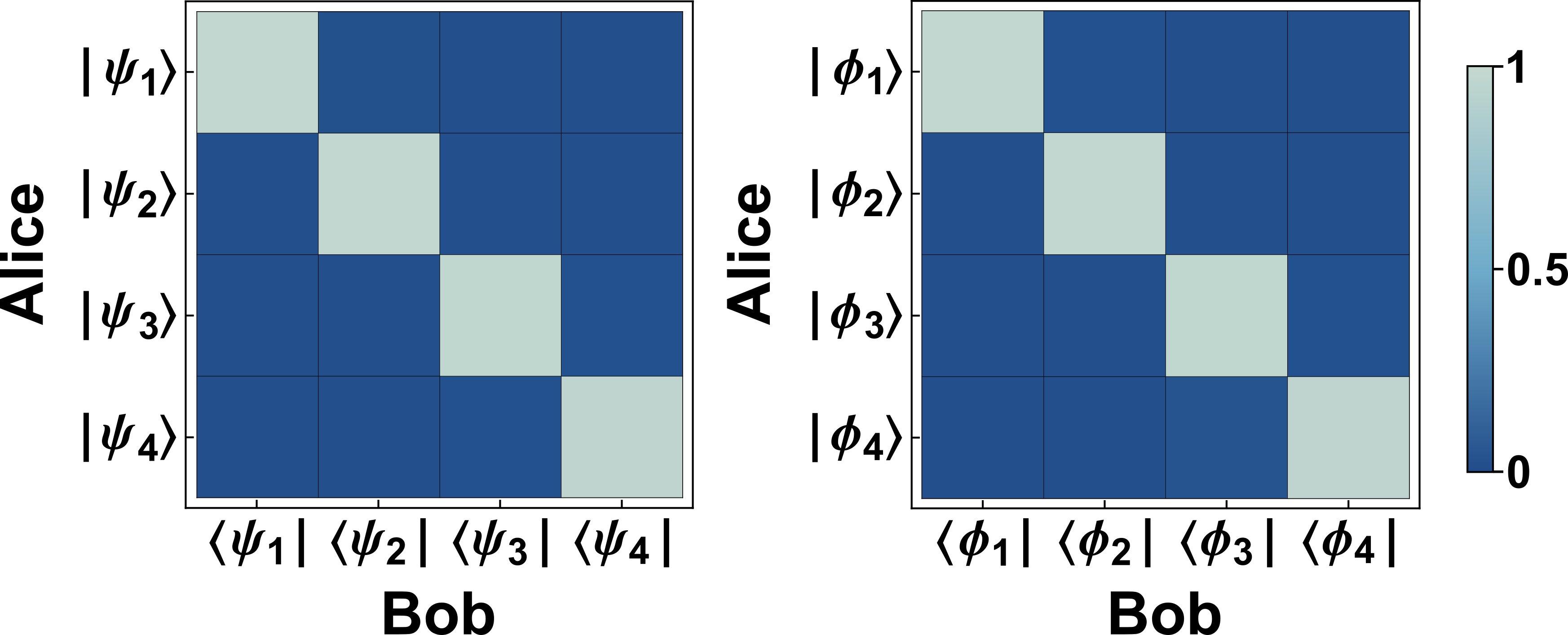}
\caption{\textbf{Probability distributions of the transmitted states.} On the left, we show the probability distribution of the $\ket{\psi_i}$ states, and on the right the $\ket{\phi_i}$ states. Data are acquired for 5 minutes, when an average photon number per pulse of $\nu=0.0026$ photon/pulse was sent.}
\label{fig_probdist}
\end{figure}

The counter-propagating control signal is sent to the receiver side where it is further attenuated such that its leakage to the quantum channel is reduced. Indeed, the circulators we used have a finite extinction ratio of 55 dB the second and 53 dB the first. We measured that with a launch power of approximately -81 dBm (\textit{i.e.} at the input of the second circulator) the leakage was negligible, \textit{i.e.} hidden by the detectors dark counts. Then, the control signal is injected in the system through the second circulator, split by a beam splitter into the two cores to stabilize, interfered on the beam splitter at transmitter's side and directed thanks to the first circulator to detector D5 (10\% efficiency and 5 $\mu$s dead time). The counts from this detectors are monitored by the phase-locked loop that drives the phase shifter in the system.

Fig.~\ref{fig_probdist} reports the probability distributions for the measured states in the two basis. The mean photon number per pulse was $\nu=0.0026$ photon/pulse, a low enough value to avoid saturation regime effects. Each acquisition was carried out for 5 minutes, and the average results are reported. Every state was successfully detected with an average fidelity of (97.73$\pm$0.04)\% and (97.57$\pm$0.04)\% for the transmission of bases $\mathcal{M}_0$ and $\mathcal{M}_1$ respectively, with the lowest measured fidelity being (96.72$\pm$0.04)\% of state $\ket{\phi_4}$.

\begin{figure}[!ht]
\centering
\includegraphics[width=.48\textwidth]{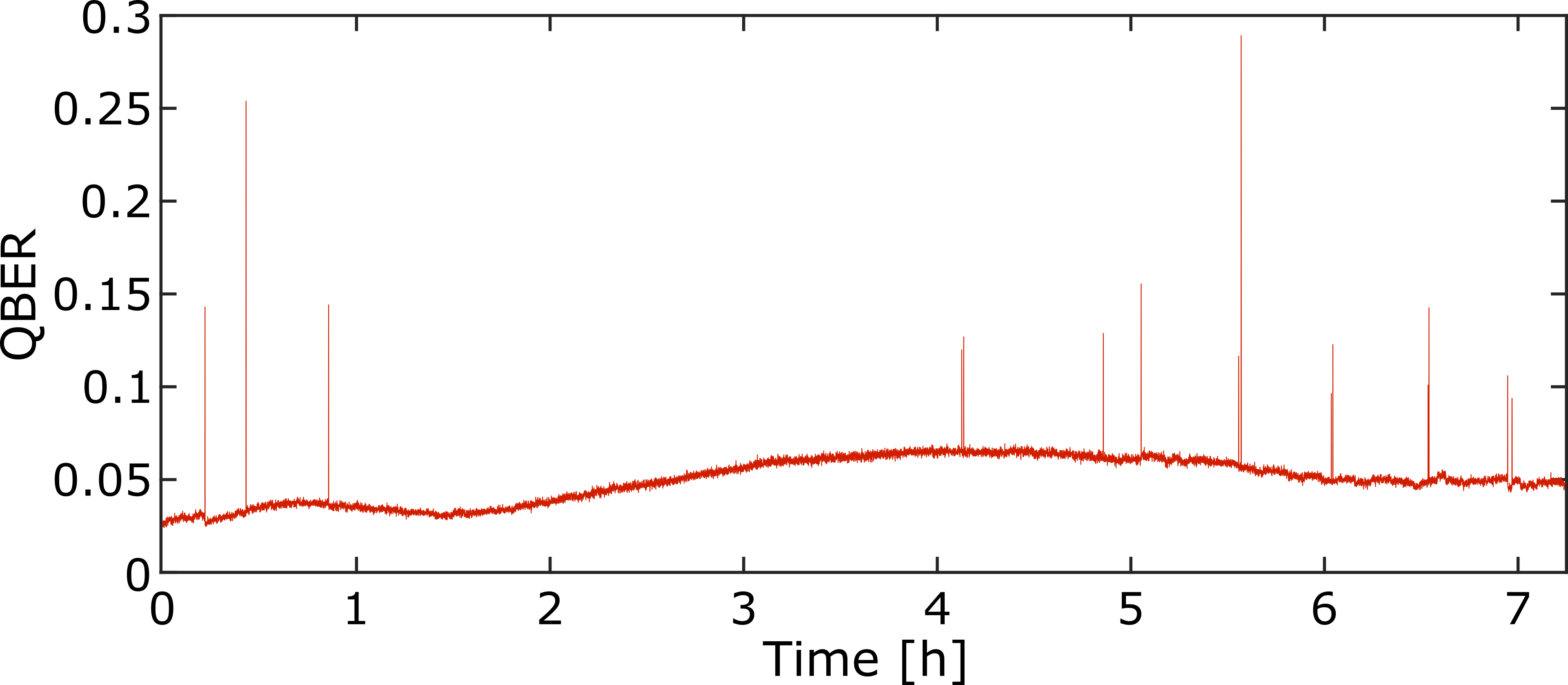}
\caption{\textbf{System stability.} The QBER of state $\ket{\phi_2}$ (with $\omega=0.0052$ photon/pulse) is monitored for 7 hours of continuous acquisition from the free-running system.}
\label{fig_stability}
\end{figure}

Moreover, we show in Fig.~\ref{fig_stability} the long term behaviour of our setup, in particular of the phase-locked loop stabilization system. We plot the quantum bit error rate (QBER) as a function of the acquisition time over a temporal window of 7 hours. During this acquisition period, the phase drifts were automatically compensated by the board. The state prepared was $\ket{\phi_2}$: the QBER then corresponds to the number of clicks collected in the other states of basis $\mathcal{M}_1$ (projections into the \textit{wrong} states) over the total number of detected clicks. {\color{black} The errors (wrong clicks) derive from experimental imperfections both during the state preparation and the channel stabilization.} The average photon number used during the acquisition was $\omega=0.0052$ photon/pulse: this value, higher than $\nu$, still allows the detectors to avoid strong saturation regimes effects. 

Where the QBER sees abrupt changes with higher values, the stabilization system lost the locking position. However, as can be seen, it is able to rapidly recover a stable interference in a position close to the one it had before. 
As a final note, we believe that the slow QBER increase as time passes is mainly due to polarization drifts, which {\color{black} have a twofold effect: on the one side, they} prevent the locking system to see a complete interference fringe and therefore to lock in the exact optimal position. {\color{black} On the other side, they also affect the visibility of the quantum state that is sent and measured.}

Finally, we implement a high-dimensional quantum key distribution protocol using the setup just described. This only provides a conceptual proof of the feasibility of such a communication protocol, as there is no real-time choice of basis nor state. Furthermore, quantum key distribution protocols require to break the phase coherence among consecutive states (to assure security against coherent attacks), which was not realized in our setup. However, we successfully prove that the stability issue can be solved by such a system to a degree where the average QBER recorded is approximately 2.5\%, much lower than those reported in the previous works, and over a much longer link length. We emulated a decoy-state technique, to prevent the photon number splitting attack against weak coherent pulses implementations, by preparing every state with two intensity levels: a signal state with $\mu_1=0.0052$ and a decoy state with $\mu_2=0.0026$ photon/pulse. Only two intensity levels are sufficient, since a one-decoy technique was shown to efficiently counteract the photon number splitting attack~\cite{rusca2018finite}. Hence, the finite key regime secret key length $l$ in such a four-dimensional scenario is given by the following equation:
\begin{multline}
    l=2s_{Z,0}^l+s_{Z,1}^l(2-h_{HD}(\phi_Z^u))-\lambda_{EC}\\
    -6\log_2(19/\epsilon_{sec})-\log_2(2/\epsilon_{cor}).
    \label{eq_skr}
\end{multline}
In eq.~\eqref{eq_skr}, $Z=\mathcal{M}_0$ is the first basis used in the protocol, $s_{Z,0}^l$ ($s_{Z,1}^l$) is the lower bound on the zero-photon (single-photon) events in the first basis, \textit{i.e.} detections due to pulses with zero (one) photons, $h_{HD}(x)=-x\log_2(x/(d-1))-(1-x)\log_2(1-x)$ is the $d$-dimensional entropy (in our experiment $d=4$) and $\phi_Z^u$ is the upper bound on the phase error rate. Error correction is not performed, but its effects are taken into consideration through $\lambda_{EC}=1.16\, n_Z\, h_{HD}(QBER_Z)$, which represents the number of bits that would be disclosed during this procedure. Here, the block size $n_Z$ and the experienced QBER in the first basis ($QBER_Z$) are used. Finally, $\epsilon_{sec}$ and $\epsilon_{cor}$ are the secrecy and correctness parameters. Formulas for each of these terms can be found in literature~\cite{pirandola2019advances,lim2014concise,rusca2018finite,islam2017provably}, and are not reported as they are out of the scope of this work. By setting the block size to $n_Z=10^{9}$, and the secrecy and correctness parameters to $\epsilon_{sec}=\epsilon_{cor}=10^{-15}$, provided with our setup channel loss, receiver loss and detectors parameters, the optimal secret key that can be generated is of approximately 5$\times10^{-5}$ bit/pulse, resulting in a rate of 29.8 kbit/s. This is obtained for: probability of sending signal and decoy states $p_{\mu_1}=0.85$ and $p_{\mu_2}=0.15$ and probability of sending the first basis $p_Z=0.91$.
To be noted that, in our setup we are just sending one state at a time, avoiding a real-time state modulation. In a real-time scenario, the intrinsic error experienced by the system could increase due to the presence of other devices (\textit{e.g.} phase modulators and optical switches), which would result in a lower key generation rate.
On the other hand, in our setup, most of the photons sent are directed to only one detector. This leads that particular detector to experience saturation regime effects for much lower mean photon number per pulse values than it would if all states were prepared and thus detected simultaneously. Thus, the maximum average photon number that we can use is bounded by much lower values, limiting the secret key rate as well.
As a final consideration, we believe that the realization of a full real-time quantum key distribution scheme exploiting our setup and the stabilization system studied in this work, would be an interesting future study. {\color{black} In particular, we believe that the choice of basis and state at Alice side should not be handled by the same system used to stabilise the channel. Indeed, the speed of the implemented stabilization board is limited to a few hundred kHz range (mainly bounded by the analog acquisition system), while the repetition rate of the system can easily reach MHz or GHz regimes. To avoid such reduced rates, we believe the two signals (quantum and control) should be completely independent, in a way that the control signal is able to simultaneously stabilize all four cores. This could be achieved by using different beam splitters for the stabilization systems and for the quantum states preparation and detection.
Moreover, a real time system would require a high-speed active choice at both Alice and Bob sides: the choice of cores can be implemented by tuning the beam splitters used for the quantum signal, while the 0 or $\pi$ phase shift can be imposed with a phase modulator. These effects can be obtained with the use of fast switches replacing the static 50/50 beam splitters and standard phase modulators, both based on the electro-optic effect of Lithium Niobate crystals.}
Moreover, the impact of such a phase-locked loop setup should also be investigated on longer multicore fiber links to fully understand its potential.

\section{Conclusion}
Quantum networks constitute one of the main goals towards a worldwide quantum internet. However, current quantum systems present limitations in terms of  transmission rate and reachable distances, thus limiting the development of this technology.\\
Here, we demonstrated how path encoded high-dimensional states, reliably transmitted through a multicore fiber, could play a prominent role in the near future quantum communication network. Indeed, taking advantage of a phase-locked loop system, which guarantees a stable interference, we transmitted high-dimensional quantum states over 2 km of multicore fiber with average fidelities of 97$\%$. Thus, by ensuring a trustworthy communication, we showed how our implementation can benefit from the advantages given by qudits, \textit{e.g.} their higher noise resilience and larger information capacity.\\
Our implementation overcome the limitations of the previous experiments, and will act as a demonstration towards the adoption of multicore fiber for distributing path encoded qudits over long distance.

\section*{Acknowledgment}
We thank OFS-Fitel for the fiber fabrication.
\bibliographystyle{IEEEtran}
\bibliography{Refs}

\begin{wrapfigure}{l}{0.1\textwidth}
  \begin{center}
    \vskip -1\baselineskip plus -1fil 
    \includegraphics[width=0.11\textwidth]{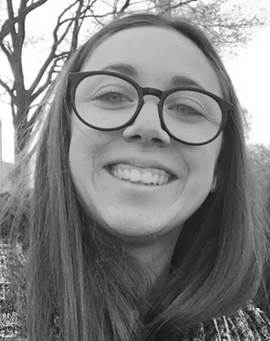}
  \end{center}
\end{wrapfigure}
\begin{IEEEbiographynophoto}{Beatrice Da Lio}
is a Ph.D. candidate of the SPOC center at the Department of Photonics Engineering at the Technical University of Denmark (DTU). She obtained her B.Sc at the University of Padova. She holds a double M.Sc degree in Engineering Telecommunication at the University of Padova and at the Technical University of Denmark. 
Her research interests are focused on quantum cryptography and quantum communication.
\end{IEEEbiographynophoto}

\vskip -1\baselineskip plus -1fil
\begin{wrapfigure}{l}{0.1\textwidth}
  \begin{center}
    \vskip -1\baselineskip plus -1fil
    \includegraphics[width=0.11\textwidth]{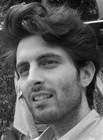}
  \end{center}
\end{wrapfigure}
\begin{IEEEbiographynophoto}{Davide Bacco}
is an Assistant Professor at the Department of Photonics Engineering at the Technical University of Denmark (DTU). He received his degree in Engineering Telecommunication in 2011 at the University of Padova, Italy. In 2015 he finished in the same University the Ph.D. degree on Science Technology and Spatial Measures (CISAS).
His research interests regard quantum communication, quantum cryptography and silicon photonics for quantum communications.
\end{IEEEbiographynophoto}

\vskip -1\baselineskip plus -1fil
\begin{wrapfigure}{l}{0.1\textwidth}
  \begin{center}
    \vskip -1\baselineskip plus -1fil
    \includegraphics[width=0.11\textwidth]{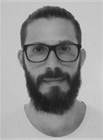}
  \end{center}
\end{wrapfigure}
\begin{IEEEbiographynophoto}{Daniele Cozzolino}
is a Ph.D. candidate of the SPOC center at the Department of Photonics Engineering at the Technical University of Denmark (DTU). He obtained his B.Sc and M.Sc in Physics at the University of Naples Federico II. 
His research interests are focused on quantum information and fundamental physics.
\end{IEEEbiographynophoto}
\vspace{0.2cm}
\begin{wrapfigure}{l}{0.11\textwidth}
  \begin{center}
    \includegraphics[width=0.11\textwidth]{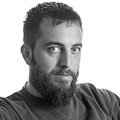}
  \end{center}
\end{wrapfigure}
\begin{IEEEbiographynophoto}{Nicola Biagi}
obtained his PhD in 2019 at the University of Florence, Italy. He is currently a Postdoc researcher at the INO-CNR, Istituto Nazionale di Ottica. His main achievements are the experimental implementation of a sequence of single photon annihilation and creation operators, in order to emulate a strong Kerr nonlinearity, and the generation and characterization of entangled macroscopic states of light. His research interests are focused on the investigation and manipulation of quantum properties of light.
\end{IEEEbiographynophoto}
\vspace{0.2cm}

\begin{wrapfigure}{l}{0.11\textwidth}
  \begin{center}
    \vskip -0.5\baselineskip plus -1fil
    \includegraphics[width=0.11\textwidth]{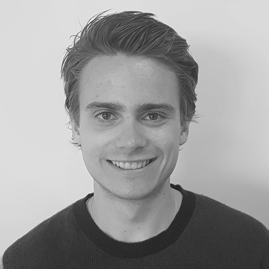}
  \end{center}
\end{wrapfigure}
\begin{IEEEbiographynophoto}{Tummas Napoleon Arge}
is a MSc. student studying Physics and Nanotechnology at the Technical University of Denmark (DTU). His interest is focused on quantum information and experimental quantum technologies.
\end{IEEEbiographynophoto}

\vskip -1\baselineskip plus -1fil
\vspace{0.2cm}
\begin{wrapfigure}{l}{0.11\textwidth}
  \begin{center}
    \vskip -1.5\baselineskip plus -1fil
    \includegraphics[width=0.11\textwidth]{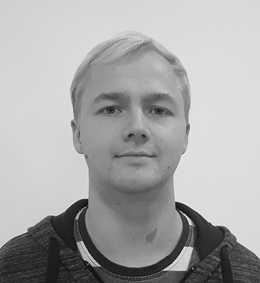}
  \end{center}
\end{wrapfigure}
\begin{IEEEbiographynophoto}{Emil Larsen}
is a MSc. student studying Photonics at the Technical University of Denmark (DTU). His interest is in theoretical optics and quantum communication. 
\end{IEEEbiographynophoto}

\vskip -0.5\baselineskip plus -1fil
\begin{wrapfigure}{l}{0.11\textwidth}
  \begin{center}
    \includegraphics[width=0.11\textwidth]{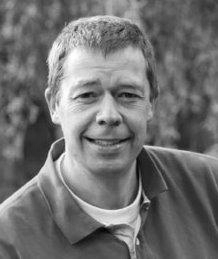}
  \end{center}
\end{wrapfigure}
\begin{IEEEbiographynophoto}{Karsten Rottwitt}
is the group leader of the Fiber Optics, Devices and Non-linear Effects group at the Department of Photonics Engineering at the Technical University of Denmark. He obtained his PhD in 1993, employed at DTU Fotonik since 2002, and full professor since 2011.  He is a member of Editorial board OPTICA and in 2019 appointed senior member OSA. His research interests are focused on nonlinear effects in optical fibers, fundamental aspects as well as applications. He has co-authored one graduate level text book on nonlinear optics.
\end{IEEEbiographynophoto}

\vskip -1\baselineskip plus -1fil
\begin{wrapfigure}{l}{0.11\textwidth}
  \begin{center}
    \includegraphics[width=0.11\textwidth]{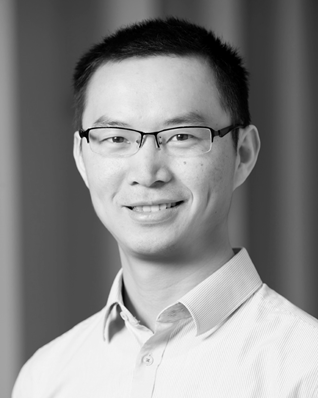}
  \end{center}
\end{wrapfigure}
\begin{IEEEbiographynophoto}{Yunhong Ding}
received the B.S. degree (with first-class honors) from Huazhong University of Science and Technology (HUST), Wuhan, China, in 2006, the Ph.D. degree in electronic science and technology from the same university in 2011. He joined the Department of Photonics Engineering (DTU Fotonik) as a postdoc in 2011. Since 2017 he has become a Senior researcher/associate professor, and was granted the Villum Young Investigator in 2019. Dr. Ding is also the founder and CEO of the spin-out company SiPhotonIC ApS. His research interests include integrated quantum information processing, quantum communication, graphene photonics, and space division multiplexing.
\end{IEEEbiographynophoto}

\vskip -1\baselineskip plus -1fil
\begin{wrapfigure}{l}{0.11\textwidth}
  \begin{center}
    \includegraphics[width=0.11\textwidth]{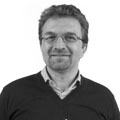}
  \end{center}
\end{wrapfigure}
\begin{IEEEbiographynophoto}{Alessandro Zavatta}
is a research scientist at the Istituto Nazionale di Ottica (INO) of the National Research Council of Italy (CNR) since 2008. He received the Laurea degree (M.Sc.) in Physics from the University of Bologna, and the Ph.D. degree from the University of Firenze (Italy). His research activities fall in the emerging field of quantum technologies, spacing from fundamental tests of quantum mechanics to the realization of innovative quantum optical devices for quantum communications.
\end{IEEEbiographynophoto}

\vspace{-1cm}

\begin{wrapfigure}{l}{0.11\textwidth}
  \begin{center}
    \includegraphics[width=0.10\textwidth]{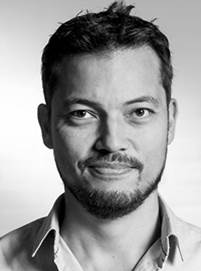}
  \end{center}
\end{wrapfigure}
\begin{IEEEbiographynophoto}{Leif Katsuo Oxenl\o we}
is the group leader of the High-Speed Optical Communications group at the Department of Photonics Engineering, at the Technical University of Denmark (DTU), and the Centre Leader of the Research Centre of Excellence SPOC (Silicon Photonics for Optical Communications) supported by the Danish National Research Foundation. He received the B.Sc. degree in physics and astronomy from the Niels Bohr Institute, University of Copenhagen, Denmark in 1996. In 1998 he received the International Diploma of Imperial College, London, UK and the M.Sc. degree from the University of Copenhagen. He received the Ph.D. degree in 2002 from DTU and since 2009 is Professor of Photonic Communication Technologies.
His research interests are focused on silicon photonics for optical processing and high-speed optical communication.
\end{IEEEbiographynophoto}

\end{document}